\documentclass[12pt]{iopart}
\usepackage{iopams}
\usepackage{epsf}
\usepackage{euscript}
\usepackage{graphicx}
\newcommand{\Cov}{\mathop{\mathrm{Cov}}\nolimits}

\newcommand{\diam}{\mathop{\mathrm{diam}}\nolimits}

\newcommand{\be}{\begin{eqnarray}}

\newcommand{\ee}{\end{eqnarray}}

\begin{document}
\title {Eigenvectors of the discrete Laplacian on regular graphs -
a statistical approach}
\author{Yehonatan Elon$^1$}
\address {$^1$Department of Physics of Complex Systems,\\
 The Weizmann Institute of Science, 76100 Rehovot, Israel}

\begin{abstract}
In an attempt to characterize the structure of eigenvectors of
random regular graphs, we investigate the correlations between the
components of the eigenvectors associated to different vertices. In
addition, we provide numerical observations, suggesting that the
eigenvectors follow a Gaussian distribution. Following this
assumption, we reconstruct some properties of the nodal structure
which were observed in numerical simulations, but were not explained
so far \cite{Dekel}. We also show that some statistical properties
of the nodal pattern cannot be described in terms of a percolation
model, as opposed to the suggested correspondence \cite{Bogomolny02}
for eigenvectors of 2 dimensional manifolds.
\end{abstract}

\section{Introduction}\label{sec:introduction}
In the past few decades, the spectral properties of regular graphs
had attracted considerable attention of researchers from diverse
disciplines such as combinatorics, information theory, theoretical
and applied computer science, quantum chaos and spectral theory (to
list only a few). In order to understand better the eigenvectors of
the Laplacian on such graphs, we try to establish some analogies
between those eigenvectors, to eigenvectors of chaotic manifolds.
The tools we are using for this task, are mostly probabilistic.\\
In the following, we consider the statistical properties of
$G(n,d)=(V,E)$ - a graph which is chosen uniformly at random from
the set of $d-$regular graphs on $n$ vertices. The graph can be
uniquely described by its \textit{adjacency matrix} $A$ (also known
as the connectivity matrix), where $A_{i,j}=1$ if $v_i$ and $v_j$
are adjacent vertices in $G$, or zero otherwise. The action of the
discrete Laplacian on a function $f:V\rightarrow\mathbb{R}$ is
 \be
  Lf_i=\sum_{i\sim j}(f_i-f_j)
 \ee
where $f_i\equiv f(v_i)$, and the summation is over all vertices
$v_j$ which are adjacent to $v_i$. For regular graphs, the Laplacian
can be expressed in a matrix form as $L=dI-A$, therefore an
eigenvector of the Adjacency matrix with an eigenvalue $\lambda$, is
also an eigenvector of the Laplacian, with an eigenvalue
$\mu=d-\lambda$.\\
The eigenvalues and eigenvectors of $A$ contain valuable information
about the structure of the graph. The relations between the spectrum
of the adjacency matrix to the expansion properties of $G$ (see
section \ref{subsec:GraphsIntro}), have been thoroughly investigated
and were found useful for coping with a variety of tasks. To mention
some, the study of expanders is related to evaluation of convergence
rates for Markov chains and the study of metric embeddings in
mathematics. In computer science, one uses expanders for the
analysis of communication networks, construction of efficient
error-correcting codes, and the theory of pseudorandomness (for a
detailed survey, see \cite{LinialExpanders}). The eigenvectors of
$A$ are being successfully used in various algorithms, such as
partitioning and clustering (e.g. \cite{Shi,Coifman,Pothen}).\\
As one can learn from spectral properties about the structure of the
graph, we can go the other way around. In the study of quantum
properties of (classically) chaotic systems, one is commonly
interested in statistical properties of the spectrum and eigenstates
of the corresponding Schr\"odinger operator. While quantum operators
on graphs (such as the Laplacian) are easy to define, the classical
analogue is not obvious. A plausible classical extension would be to
consider a random walk on the graph. For a connected graph which is
not bipartite, it is known that random walks are mixing fast (e.g.
\cite{Lovasz}). Since a fast mixing system is chaotic, one might
expect that the quantum properties of a generic graph will be
related in some manner to those of chaotic systems. This conjecture
was supported by numerical simulations \cite{Rudnick}, and recently
found an explicit formulation \cite{Smilansky07}, relating spectral
properties to cycles in $G$.\\
The main goal of the current paper is the characterization of
statistical properties of eigenvectors of $(n,d)$ graphs. As this
work is inspired by analogue findings for chaotic wave functions,
and uses extensively several combinatorial properties of random
regular graphs, we dedicate the next section to review some relevant
results concerning chaotic billiards and $(n,d)$ graphs.\\
In section 3 we examine correlations of the eigenvectors at
different vertices. We derive an explicit limiting expression for
the (short distance) empirical covariance, which depends only on the
eigenvalue $\lambda$.\\
In section 4 we provide numerical evidence which suggest that the
distribution of the eigenfunctions' components can be approximated
by a Gaussian measure.\\
Assuming a Gaussian measure, we dedicate section 5, to the
evaluation of some expected properties of the nodal pattern of the
eigenvectors, such as the expected number of nodal domains and their
expected structure.

\section{A brief review of previous results}
\subsection{Eigenvectors of chaotic billiards}\label{subsec:QC}
A classical billiard system is defined as a point particle, which is
confined to a domain $\mathcal{D}\subset\mathbb{R}^n$. The particle
moves with a constant speed along geodesics and collides specularly
with the boundary of $\mathcal{D}$. Depending on the shape of the
boundary, the dynamics of the particle can be classified as chaotic
or regular. A quantum analogue would be to consider eigenstates of
the Schr\"odinger operator for a particle confined to $\mathcal{D}$:
 \be\label{eq:LaplaceBeltrami}
  -\Delta_D\psi(\mathbf{r})=k^2\psi(\mathbf{r})
 \ee
- the Laplace-Beltrami operator,restricted to $\mathcal{D}$, with
Dirichlet boundary condition.\\
The statistics of the wave function $\psi(\mathbf{r})$, rely on the
classical properties of $\mathcal{D}$. In \cite{Berry77}, a limiting
expressions for the auto correlation function
 \be\label{eq:CovBerry}
  C(\mathbf{r}_0,\mathbf{r},k)=\frac
  {\langle\psi(\mathbf{r}_0+\mathbf{r}/2)\psi^*(\mathbf{r}_0-\mathbf{r}/2)\rangle}
  {\langle|\psi(\mathbf{r})|^2\rangle}
 \ee
was calculated, where $\langle...\rangle$ denotes averaging over an
appropriate spectral window $[k,k+\epsilon k]$, in the
semi-classical limit $k\rightarrow\infty$
 \footnote{As the density of states is scaled as $k^{n-1}$, we demand that $\epsilon
  k\rightarrow0$, but $\epsilon k^{n-1}\rightarrow\infty$, so that the
  energy does not vary significantly along the window, and the number of
  states is large enough.}.
It was shown that for an integrable domain,
$C(\mathbf{r}_0,\mathbf{r},k)$ is anisotropic and depend on the
symmetries of the domain. For chaotic domains, the limit of the auto
correlation function is isotropic and universal (for points which
are far enough from the boundary \cite{Urbina}), and can be written
explicitly as
 \be\label{eq:CovBerry2}
  \lim_{k\rightarrow\infty}\mathcal{C}(\mathbf{r},k)=
  \frac{\Gamma(n/2)J_{n/2-1}(|k\mathbf{r}|)}{(|k\mathbf{r}|)^{n/2-1}}
 \ee
where $J_\nu(x)$ is the $\nu$th Bessel function of the first kind,
which decays asymptotically as
$J_\nu(x)\sim\cos(x+\phi_\nu)/\sqrt{x}$. Moreover, it was suggested
that in the semi-classical limit, the eigenvectors statistics
reproduce a Gaussian measure (the \textit{random wave model}), i.e.
for $\{\mathbf{r}_1,\mathbf{r}_2...\mathbf{r}_m\}\in\mathcal{D}$,
the probability density of
$\psi(\mathbf{R})\equiv(\psi(\mathbf{r}_1),...\psi(\mathbf{r}_m))^T$
for a wave function chosen uniformly from the spectral window
$[k,k+\epsilon k]$, is converging to
 \be\label{eq:DensityBerry}
  p(\psi(\mathbf{R}))=\frac1{(2\pi)^{m/2}\sqrt{|{C_k}|}}
  \exp\left[-\frac12 \psi(\mathbf{R})^T C_k^{-1} \psi(\mathbf{R})\right]
 \ee
where the covariance matrix is given by
$(C_k)_{ij}=\lim_{k\rightarrow\infty}\mathcal{C}(\mathbf{r}_i-\mathbf{r}_j,k)$.
Although the random wave model is not supported by any rigorous
derivation, it was found consistent with some numerical
observations, such as
\cite{Blum,Elon}.\\
A different characterization of the eigenvectors, based on their
nodal pattern, was suggested in \cite{Blum} for two-dimensional
manifolds: since it is always possible to find a basis in which the
eigenfunctions of $-\Delta$ are real, they can be divided into
\textit{nodal domains}, connected regions of the same sign,
separated by \textit{nodal lines} on which the eigenfunction
vanishes. By Courant theorem \cite{Courant}, the $j$th eigenstate
contains no more than $j$ nodal domains. The authors have
investigated the limiting distribution (as $j\rightarrow\infty$) of
the parameter $\xi_j=\nu_j/j$, where $\nu_j$ is the number of nodal
domains in the $j$th wave function. They have derived an explicit
expression for separable domains, which depends on the explicit
structure of the domain. For chaotic billiards, they have observed a
universal limiting distribution, independent of the investigated
domain.\\
 This limiting distribution found an intriguing explanation by
\cite{Bogomolny02}, where the nodal pattern is described in terms of
critical bond percolation model. While for some measures on the
nodal lines \cite{Foltin04,Aronovitch} the correspondence is not
complete, general arguments such as \cite{Bogomolny07} implies that
the scaling limit of both of the models should converge. In
addition, the model predicts with a great accuracy diverse
properties of the nodal pattern and the nodal lines
\cite{Bogomolny02,BogomolnySle,Keating,Elon}.
\subsection{Some properties of large regular graphs}
\label{subsec:GraphsIntro} Throughout this paper, we will focus our
attention on $(n,d)$ graphs, where $d\ge3$ is fixed, and
$n\rightarrow\infty$. With a high probability, those graphs are
highly connected, or \textit{expanding}. An expander graph $G=(V,E)$
has the property that for every (small enough) subset $S\subset V$,
the edge boundary $\partial S$, which is the set of edges connecting
$S$ to
$G\setminus S$, is proportional in size to $S$ itself.\\
A related property of $(n,d)$ graphs, which will be used repeatedly
in the following, is the \textbf{local tree property}. It is known
\cite{Bollobas80} that for $k\le\log_{d-1}(n)$, the numbers $C_k$ of
cycles of length $k$ in an $(n,d)$ graph, are distributed
asymptotically as independent Poisson random variables with mean
$E(C_k)=(d-1)^k/2k$. Therefore, for any $\epsilon>0$ and as
$n\rightarrow\infty$, almost all of the vertices of an $(n,d)$ graph
are not contained in a cycle of shorter length than
$(1-\epsilon)\log_{d-1}(n)$, with high probability. Equivalently,
the ball of radius $\log_{d-1}(n)/(2+\epsilon)$ around almost all of
the vertices is a tree. The volume of a ball of radius $k$ in an
$(n,d)$ graph grows exponentially with $k$ for $k\le\log_{d-1}(n)$.
In fact, the diameter of $G$ may differ from $\log_{d-1}(n\log n)$
only by a (small) finite number independent of $n$ \cite{Bollobas}.
In the following we will express logarithms in the natural tree
base:
$\log(x)\equiv\log_{d-1}(x)$.\\
The adjacency matrix of a graph is real and symmetric, therefore it
has a real spectrum, which is supported on $[-d,d]$. As
$n\rightarrow\infty$, the spectral measure on $A$ converges to the
Kesten-McKay measure \cite{McKay}:
 \be\label{eq:KestenMcKay}
  p(\lambda)=\left\{\begin{array}{lcr}\frac{d}{2\pi}\frac{\sqrt{4(d-1)-\lambda^2}}{d^2-\lambda^2}
  & \mbox{for} & |\lambda|\le2\sqrt{d-1}\\
  0 & \mbox{for} & |\lambda|>2\sqrt{d-1}\end{array}\right.
 \ee
We will use the following notation throughout this paper:
Eigenvalues of the adjacency matrix are denoted by $\lambda$ and
those of the Laplacian by $\mu$; superscript indices denote
eigenvectors: $Af^{(i)}=\lambda_i f^{(i)}$; Subscript indices will
denote vertices: $f^{(i)}_j=f^{(i)}(v_j)$. We choose the
normalization $\langle f,f\rangle=n$, so that $E(f_i^2)=1$,
irrespective of $n$. We enumerate the eigenvalues in the customary
order: $d=\lambda_1\ge\lambda_2...\ge\lambda_n$, or equivalently
$0=\mu_1\le\mu_2...\le\mu_n$. The first eigenvector (or the ground
state of $L$) is the constant vector $f^{(1)}=(1,1...,1)^T$. As the
eigenvectors are orthogonal, we get for $i>1$ that $\sum_j
f^{(i)}_j=\langle f^{(1)},f^{(i)}\rangle=0$. We would like to
emphasize again that for a regular graph, the eigenvectors of the
adjacency matrix and the Laplacian are identical. Therefore, all the
results that will be derived in the following are applicable (up to
rescaling of the eigenvalue) to both of the operators.\\
For a graph $G=(V,E)$ and a function $f(V)$, a positive (negative)
nodal domain of $f$ is a maximal connected component of $G$, so that
$f(v)\ge0$ ( $f(v)\le0$ ) for all of the vertices in the component.
\footnote{In the following we will ignore the possibility that for
some vertex $f(v)$ vanishes, as this event is of measure zero for
Laplacian eigenvectors of $(n,d)$ graphs.} The nodal count of $f$,
which will be denoted by $\nu$, is the number of nodal domains of $f$.\\
In \cite{Davies}, Courant theorem is generalized to connected
discrete graphs, showing that the $j$th eigenvector of the Laplacian
contains no more than $j$ nodal domains. A constraint on the allowed
shapes of domains was derived in \cite{Band}: Since an adjacency
eigenfunction satisfies: $\lambda f_i=\sum_{i\sim j}f_j$ , if
$\lambda>k$ (for $k\in\mathbb{N}$), then for every positive
(negative) nodal domain, the maximum (minimum) of the domain must
have at least $k+1$ adjacent vertices of the same sign, therefore
the minimal size of a domain is $k+2$. Similarly, if $\lambda<0$ ,
every vertex has at least one adjacent vertex with an opposite sign,
therefore for a negative eigenvalue, nodal domains cannot have inner
vertices. In addition,by adding assumptions on the structure of the
graph (for example, by considering trees only), it is possible to
bound the minimal size of a domain for a given eigenvalue
\cite{Leydold}. We refer the reader to \cite{Biyikouglu} for a
review on the nodal pattern of general graphs.\\

\section{The covariance of an eigenvector}\label{sec:covariance}
In this section we would like to estimate the correlation between
two distinct components of an adjacency eigenvector in an $(n,d)$
graph. The distance in $G$ between two vertices $v_i,v_j\in V$ is
the length of the shortest walk in $G$ from $v_i$ to $v_j$ - we
denote the distance by $|i-j|$. Setting the
\textbf{\textit{k}-adjacency operator} to be
 \be\label{eq:AdjKMat}
  (\tilde{A}_k)_{ij}=\left\{\begin{array}{cl}
  1&\mbox{for }|i-j|=k\\ 0 & \mbox{otherwise}
  \end{array}\right. \ ,
 \ee
we evaluate the correlations between two components of an
eigenvector $f$ at distance $k$, by computing the \textbf{empirical
\textit{k}-covariance} of $f$ and $G$, defined as
 \be\label{eq:covDef}
  \Cov_k^{emp}(f,G)=\frac1{\mathcal{M}_k}\sum_{|i-j|=k}f_if_j
  =\frac1{\mathcal{M}_k} \langle f,\tilde{A}_k f\rangle
 \ee
where $\mathcal{M}_k=\sum_{i,j}(\tilde{A}_k)_{ij}$ is the number of
(directed) $k-$neighbors in $G$.\\
For $k<\log n/2$ , we can take advantage of the local tree property,
in order to find an explicit limiting expression for
(\ref{eq:covDef}). Under the tree approximation, $\mathcal{M}_k=
nd(d-1)^{k-1}$. Moreover, for a tree, $(\tilde{A}_k)_{ij}=1$ if and
only if there is a (unique) walk of length $k$ from $v_i$ to $v_j$
which do not retrace itself (do not backscatter) at any step.
Therefore, for a tree the operator $\tilde{A}_k$ is identical to the
'non-retracing operator', introduced and calculated in
\cite{Alon06}. Clearly, $\tilde{A}_0=I,\tilde{A}_1=A$, where $I$ is
the identity matrix. $\tilde{A}_2=A^2-dI$,  as one has to eliminate
from $A^2$ (which correspond to all possible walks of length $2$ in
$G$) the walks which return back to their origin at the second step.
In a similar manner, one gets for $k>2$ that
 \be\label{eq:AdjK}
  \tilde{A}_k=A\tilde{A}_{k-1}-(d-1)\tilde{A}_{k-2}\quad .
 \ee
The first term is due to all paths of length $k$ which do not
retrace in the first $k-1$ steps, while the second term eliminates
paths which have not retraced in the first $k-1$ steps, but do
retrace in the $k$th step. Since $A^k f=\lambda^k f$, we get by
substituting (\ref{eq:AdjK}) in (\ref{eq:covDef}) that in the limit
the empirical covariance converges to
 \be\label{eq:CovExp1}
  \Cov_k^{tree}(\lambda)=\frac1{d(d-1)^{k-1}}P_k(\lambda) \quad ,
 \ee
where $P_k(\lambda)$ is given by the recursion relation:
 \be\label{eq:Pk}
  \left\{\begin{array}{l} P_1(\lambda)=\lambda\\
   P_2(\lambda)=\lambda^2-d\\
   P_{k+2}(\lambda)=\lambda P_{k+1}(\lambda)-(d-1)P_k(\lambda) \quad .
   \end{array}\right.
 \ee
Introducing Chebyshev polynomials of the second kind
\cite{Abramowitz}:
 \be\label{eq:Chebyshev2}
  U_k(\cos\theta)=\frac{\sin((k+1)\theta)}{\sin\theta}
 \ee
The solution to this recursion relation, subject to the initial
conditions, can be written as
 \be\label{eq:CovExp2}
  \fl\quad\Cov_k^{tree}(\lambda)=\frac1{d(d-1)^{k/2}}
  \left((d-1)U_k\left(\frac{\lambda}{2\sqrt{d-1}}\right)-
  U_{k-2}\left(\frac{\lambda}{2\sqrt{d-1}}\right)\right)\quad .
 \ee
The functions $\Cov_k^{tree}(\lambda)$ are orthogonal polynomials of
degree $k$ in $\lambda$ with respect to Kesten-McKay measure
$p(\lambda)$ (\ref{eq:KestenMcKay}), satisfying
   \be\label{eq:Orthogonal}
    \int
    \Cov_k^{tree}(\lambda)\Cov_{k'}^{tree}(\lambda)p(\lambda)d\lambda
    =\frac1{d(d-1)^{k-1}}\delta_{k,k'}\quad .
   \ee
This results has a simple combinatorial interpretation. Following
(\ref{eq:covDef}), the left hand side of (\ref{eq:Orthogonal}) is
nothing but
 \be\label{eq:Trace}
  \lim_{n\rightarrow\infty} \frac{n}{\mathcal{M}_k\mathcal{M}_{k'}}
  \Tr(\tilde{A}_k\tilde{A}_{k'})=
   \frac1{d(d-1)^{k-1}}\frac1{\mathcal{M}_{k'}}\Tr(\tilde{A}_k\tilde{A}_{k'})
 \ee
Note that $\Tr(\tilde{A}_k\tilde{A}_{k'})$ is the number of closed
walks in $G$, which are combined from a non retracing walk of length
$k$, followed by a non retracing walk of length $k'$. The only way
to perform such a walk on a tree is by going back and forth,
therefore $\Tr(\tilde{A}_k\tilde{A}_{k'})=0$, if $k\ne k'$, or
$\mathcal{M}_{k}$ (the number of non backscattering walks of length
$k$) for $k=k'$. A substitution yield the identity
(\ref{eq:Orthogonal}).\\
As $|U_k(x)|\le k$ for $|x|\le 1$, the limiting expression for the
covariance (\ref{eq:CovExp2}) is an oscillatory function which
decays as  $(d-1)^{-k/2}$. This behavior is analogous to the
expected rate of decay for continuous chaotic manifolds
(\ref{eq:CovBerry2}). The surface of a ball of radius $r$ in
$\mathbb{R}^n$ grows as $S(r)\sim r^{n-1}$, while for regular trees,
the surface of the ball grows as $S(r)\sim(d-1)^r$. Therefore, in
both of the cases, the rate of decay of the covariance is
proportional to the root of the area of the sphere.

\begin{figure}[h]
\centering
 \scalebox{0.6}{\includegraphics{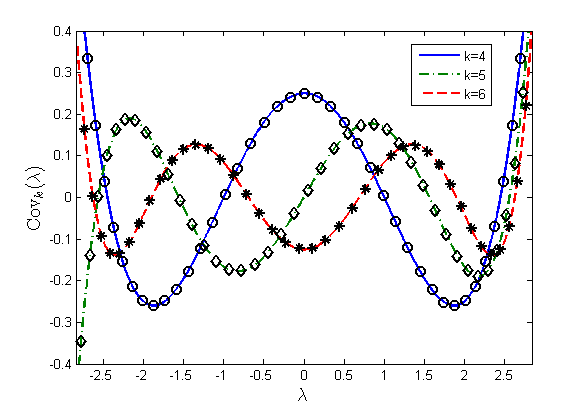}}
 \scalebox{0.6}{\includegraphics{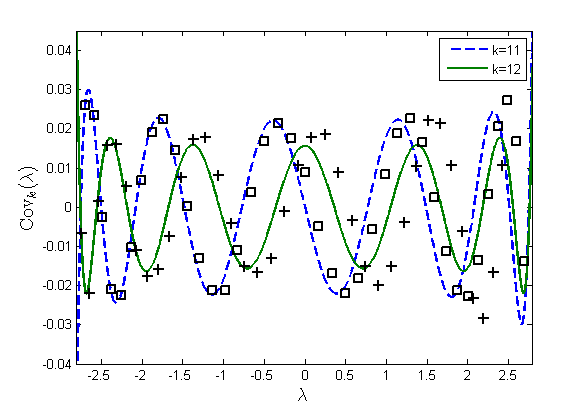}}
  \caption{A comparison between $\Cov_k^{tree}(\lambda)$ (marked by lines), and
  $\Cov_k^{emp}(f,G)$, for a single realization of $G(4000,3)$ (denoted by
  different markers), where $\log_{d-1}(n)=11.97$.\\
  Upper figure: a comparison for $k=4,5,6$. Lower figure: $k=11,12$.}
    \label{fig:3correlations}
\end{figure}

\begin{figure}[h]
\centering
 \scalebox{0.5}{\includegraphics{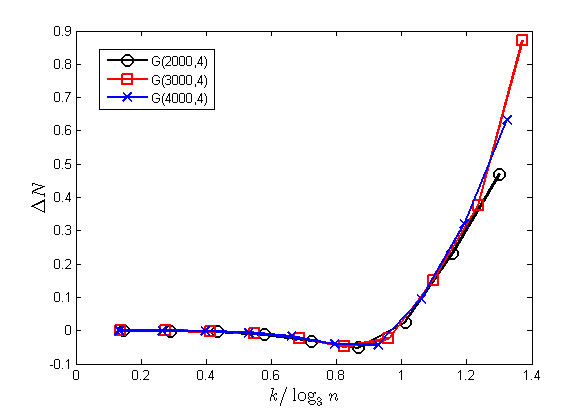}}
 \scalebox{0.5}{\includegraphics{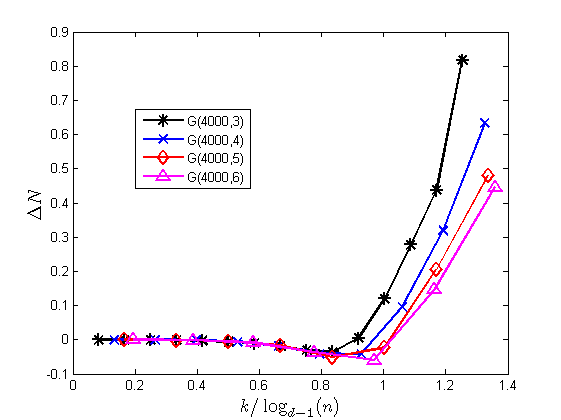}}
  \caption{\label{full} The scaled norm deviation (\ref{eq:NormDev}) for
  several realizations of regular graphs.\\
  Left: $\Delta N$, as a function of $k/\log_3(n)$, for 3 realizations of
  4-regular graphs consisting of 2000,3000 and 4000 vertices.\\
  Right: $\Delta N$, as a function of $k/\log_{d-1}(n)$, for $(4000,3),(4000,4)
  ,(4000,5)$ and $(4000,6)$ graphs.}
  \label{fig:ndev}
\end{figure}

While for short distances the empirical covariance converges to
$\Cov_k^{tree}(\lambda)$, the validity of this approximation is
expected to deteriorate  as $k$ exceed $\log n/2$. As the
computation presented above does not provide an error estimate, we
have turned to numerical simulations. We have calculated numerically
the empirical covariance (\ref{eq:covDef}) for several realizations
of regular graphs, and compared them to the limiting expression
(\ref{eq:CovExp2}). The graphs were generated following
\cite{Steger} and using \textit{MATLAB}. As expected, for short
distances the matching is very good, while for $k\sim\log n$ the
deviations are evident. Figure \ref{fig:3correlations} demonstrate
this behavior for a realization of $(4000,3)$ graph, where
$\log_2(4000)=11.97$.\\
\noindent Although for large distances, the empirical covariance
deviates from $\Cov_k^{tree}(\lambda)$, it seems that the expected
rate of decay is reproduced quite well, so that asymptotically
$|\Cov_k^{emp}(f,G)|\sim(d-1)^{-k/2}$. In order to test this
assumption, we introduce for a given $G(n,d)$ and $k$, the two norms
 \be\label{eq:NormDef}
  N_k^{emp}(G) =\frac1n\sqrt{\sum_n(\Cov_k^{emp}(f^{(i)},G))^2}\\\nonumber
  N_k^{tree}(G)=\frac1n\sqrt{\sum_n(\Cov_k^{tree}(\lambda_i))^2}
 \ee
and define the (scaled) norm deviation as:
 \be\label{eq:NormDev}
  \Delta N(G,k)=\frac{N_k^{emp}-N_k^{tree}}{N_k^{emp}+N_k^{tree}}
 \ee
If the asymptotic rate of decay of $\Cov_k^{tree}$ and
$\Cov_k^{emp}$ is similar \footnote{meaning that for some positive
$c_1(d),c_2(d)$ and for all $n$ and $k$, we get with high
probability that $c_1 N_k^{tree}(G)\le N_k^{emp}(G)\le c_2
N_k^{emp}(G)$.}, then $|\Delta N|$ will be bounded away from 1 for
all $k$ and $n$.\\
The (average) norm deviation of an eigenvector is a function of the
three parameters $n,d$ and $k$. However, a comparison of the norm
deviation for several realizations of $(n,d)$ graphs with various
values of $n$ and $d$, suggest that it might be well approximated by
a function of two parameters only - $d$, and the scaled parameter
$k/\log_{d-1}(n)$ (see figure \ref{fig:ndev}-left). For a fixed
value of $\log_{d-1}(n)$, the deviation decreases as $d$ increases
(figure \ref{fig:ndev}-right). In addition, for $k/\log n\le1$, the
norm deviation is bounded away from one, for any $d$. Since the
diameter of an $(n,d)$ graph is very close to $\log(n\log n)$, we
get that $\max(k)/\log n$ approaches one as $n$ approaches infinity,
therefore we expect $|\Delta N|$ to be bounded away from one for all
$k$.

\section{The limiting distribution of eigenvectors}\label{sec:normalDist}
In the last section we have shown some of the similarities between
the limiting expressions for the covariance of regular graphs and
the autocorrelation of chaotic billiards $(\ref{eq:CovBerry2})$. In
this section and the next, we present extensive numerical evidence,
which suggest that the distribution of eigenvectors of $(n,d)$
graphs follows a Gaussian measure, resembling the conjectured
distribution \cite{Berry77} for eigenvectors of quantum billiards
(see section
\ref{subsec:QC}).\\
In order to examine the limiting distribution of the eigenvectors
components of an $(n,d)$ graph, we have to define at first what is
the ensemble we are interested in. As an example, we can fix a graph
$G=(V,E)$, a vertex $v_i\in V$ and ask for the distribution
$f^{(j)}_i$ of the $i$th component of a randomly chosen eigenvector.
A second option is to fix an eigenvalue $\lambda_0\in [-2\sqrt{d-1},
2\sqrt{d-1}]$, and ask for the limiting distribution of an arbitrary
vertex, where we choose a graph on random, and look at the
eigenvector which has the closest eigenvalue to $\lambda_0$. In the
same manner, it is possible to fix an $(n,d)$ graph, an eigenvector
$f^{(j)}$ and ask for the limiting distribution
$f^{(j)}_i$ of a randomly chosen vertex $v_i\in V$.\\
In the following, we suggest that as $n\rightarrow\infty$, the
distribution of the eigenvectors components, with respect to the
first two ensembles is converging to a Gaussian. As for the third
ensemble, numerical simulations (e.g. \cite{LinialExpanders}),
together with the results of this work imply that a limiting
distribution for that ensemble may not exist.\\
For the sake of clarity, we begin by considering the limiting
distribution of a single vertex, which will be followed by the study
of a multivariate version. We will denote by $p(x)$ and $\Phi(x)$,
the density and the cumulative distribution function (cdf) of the
standard normal variable:
 \be\label{eq:NormalDist}
  p(x)=\frac1{\sqrt{2\pi}}e^{-x^2/2} \quad ,\quad
  \Phi(x)=\int_{-\infty}^x p(y)dy
 \ee
\subsection{The limiting univariate distribution}\label{subsec:univariate}
Based on numerical simulations we suggest the following limits:
 \begin{itemize}
 \item \textbf{Hypothesis I} (univariate): For asymptotically almost any $G(n,d)=(V,E)$ and $v_i\in
  V$, the probability that $f^{(j)}(v_i)<x$, where $f^{(j)}$ is an
  adjacency eigenvector, chosen uniformly from
  $\{f^{(2)},f^{(3)},...f^{(n)}\}$, is bounded by
   \be\label{eq:NormalBound}
    \left|\mathbb{P}(f^{j}(v_i)<x)-\Phi(x)\right|<\Delta(n)
   \ee
  where $\Delta(n)\rightarrow0$ as $n\rightarrow\infty$.
 \item \textbf{Hypothesis II} (univariate): For a given $\lambda_0\in[-2\sqrt{d-1},
  2\sqrt{d-1}]$ and $1\le i\le n$, the distribution of
  $f^{(\lambda_0)}(v_i)$, is converging to the normal distribution
  (\ref{eq:NormalDist}), where $f^{(\lambda_0)}$ is an adjacency
  eigenvector with the closest eigenvalue to $\lambda_0$ of a
  uniformly randomly chosen $(n,d)$ graph.
\end{itemize}

\noindent These assumptions may be examined by comparing the
appropriate empirical cumulative distribution functions to
$\Phi(x)$. A plausible measure to the distance between two
cumulative distributions $\mathcal{P}_1(x),\mathcal{P}_2(x)$ is the
Kolmogorov-Smirnov (KS) distance:
 \be\label{eq:KSstat}
  \mathcal{D}^{KS}_i=\sup\left|\mathcal{P}_1(x)-\mathcal{P}_2(x)\right|
 \ee
According to Kolmogorov theorem, if $\mathcal{P}_2(x)$ is an
empirical cdf of $n$ iid variables, generated with respect to the
cdf $\mathcal{P}_1(x)$, then the distribution of
$\mathcal{D}^{KS}_i$ is given by
 \be\label{eq:KSdist}
  \lim_{n\rightarrow\infty}\mathbb{P}\left(\mathcal{D}^{KS}_i\le
  \frac{y}{\sqrt{n}}\right)=
  1-2\sum_{q=1}^\infty (-1)^{q-1}e^{-2q^2y^2}
 \ee
\begin{figure}[h]
\centering
 \scalebox{0.5}{\includegraphics{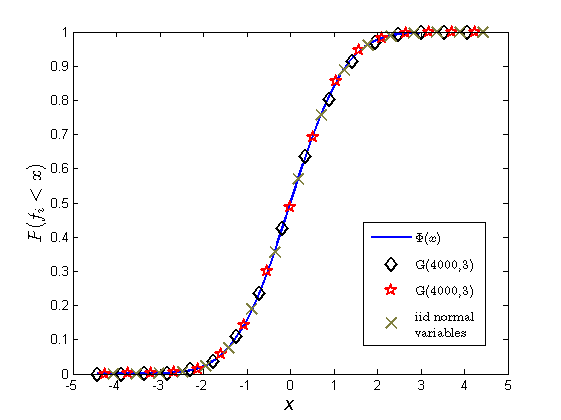}}
 \scalebox{0.5}{\includegraphics{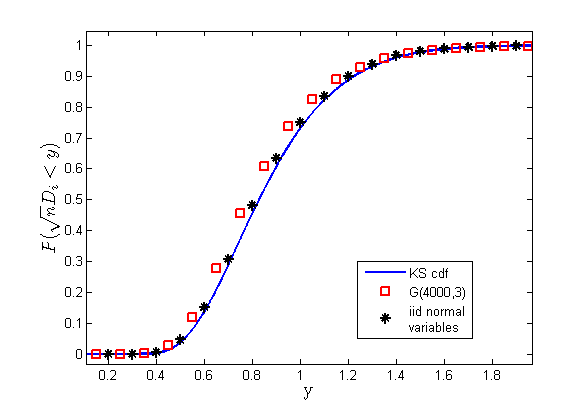}}
 \scalebox{0.5}{\includegraphics{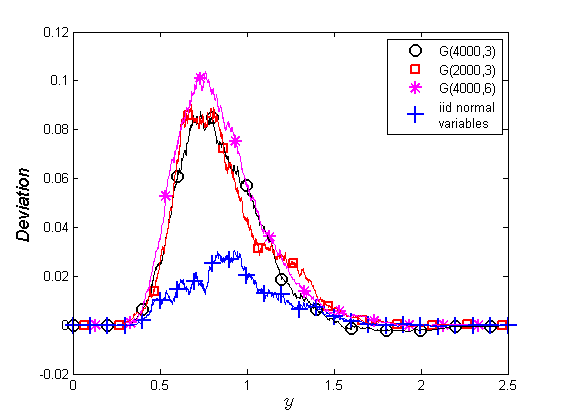}}
  \caption{Upper left figure: a comparison between $\mathcal{F}_i(x)$
  for two vertices of a realization of $(4000,3)$ graph, the empirical
  cdf of 4000 iid normal variables and $\Phi(x)$.\\
  Upper right figure: a comparison between (\ref{eq:KSdist}), the empirical
  cdf of $\mathcal{D}^{KS}$ for 4000 vertices of a $(4000,3)$ graph and the
  empirical cdf of $\mathcal{D}^{KS}$ for 4000 independent vectors of 4000
  iid normal variables.\\
  Lower figure: deviations of the empirical cdf of
  $\mathcal{D}^{KS}$ from (\ref{eq:KSdist}), for a single realization of
  $G(4000,3)$, $G(2000,3)$ , $G(4000,6)$ and the iid normal
  variables. A positive deviation corresponds to a faster convergence
  than predicted by (\ref{eq:KSdist}).}
    \label{fig:KSTest}
\end{figure}

\noindent In order to test the first hypothesis we have generated
realizations of $(n,d)$ graphs for several values of $n$ and $d$.
For a given graph and a vertex $v_i\in V$, the empirical cdf is
given by
 \be\label{eq:EmpComDis}
  \mathcal{F}_i(x)=\frac1n\#\{j|f^{(j)}_i<x\}
 \ee
A numerical comparison between $\mathcal{F}_i(x)$ to $\Phi(x)$ shows
persuasively that the differences between the two distributions are
of order $1/\sqrt{n}$, as is demonstrated in the upper left plot in
figure \ref{fig:KSTest}.\\
Since the components $f^{(j)}_i$ are not independent, the KS
distance between $\mathcal{F}_i(x)$ and $\Phi(x)$ is not expected a
priori to follow (\ref{eq:KSdist}). However, the measured KS
distances for different vertices of the same graph, was found to be
very close to (\ref{eq:KSdist}), as can be seen in figure
\ref{fig:KSTest} (upper right figure). In fact, the observed
convergence of $\mathcal{D}^{KS}$ seems to be slightly faster than
predicted by (\ref{eq:KSdist}), irrespective of $n$ and $d$ (figure
\ref{fig:KSTest}-lower figure).\\
\begin{figure}[h]
\centering
 \scalebox{0.6}{\includegraphics{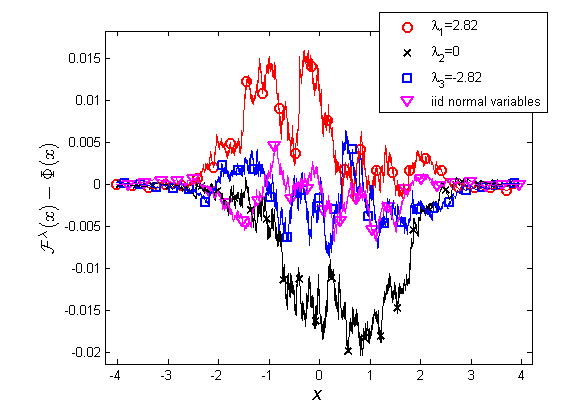}}
  \caption{The deviation of the empirical cdf, calculated according
  to the requirements of hypothesis II, from
  $\Phi(x)$. The different curves correspond to
  $\lambda_1=2.82$, $\lambda_2=0$, $\lambda_3=-2.82$
  and the empirical cdf of 10000 iid normal variables.}
    \label{fig:CdfDev}
\end{figure}
\noindent In order to examine hypothesis II, we have generated 10000
realizations of $(6000,3)$ graphs, for which the spectrum is
supported on $\pm2\sqrt{2}\approx2.828$. We have compared
$\mathcal{F}^{\lambda}(x)$ - the empirical cdf of the appropriate
components, for various values of $\lambda$, varying from $-2.82$ up
to $2.82$. In this case as well, the distance between
$\mathcal{F}^{\lambda}(x)$ and $\Phi(x)$ tends to zero as
$1/\sqrt{n}$. In addition, no substantial differences in the
limiting distribution, nor in the rate of convergence were observed
for different values of $\lambda$, as we demonstrate in figure
\ref{fig:CdfDev}.

\subsection{The multivariate limiting distribution}\label{subsec:multivariate}
\begin{figure}[h]
\centering
 \scalebox{0.6}{\includegraphics{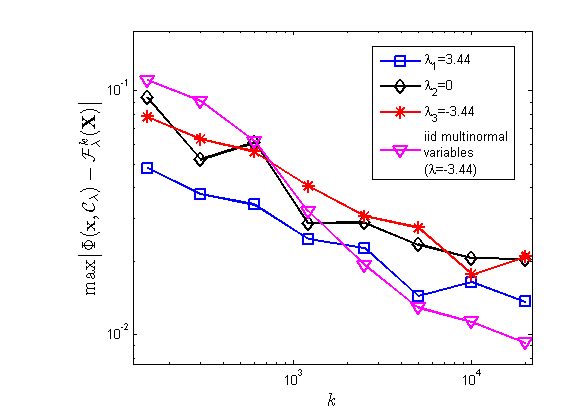}}
  \caption{The maximal deviation between the cumulative distributions
  $\mathcal{F}_\lambda^k(\mathbf{x})$ and $\Phi(\mathbf{x},\mathcal{C}_\lambda)$ as a
  function of $k$ - the number of generated graphs (logarithmic scale).
  The different curves correspond to $\lambda_1=3.44$, $\lambda_2=0$ and
  $\lambda_3=-3.44$. The maximal deviation is compared to that of $k$
  iid vectors, generated with respect to the measure $\Phi(\mathbf{x},\mathcal{C}_{-3.44})$.}
    \label{fig:CheckTwo}
\end{figure}

In order to formulate a multivariate version of the suggested
distributions, we introduce the following notation:\\
For a graph $G=(V,E)$, and $U=\{u_1,...u_m\}\subset V$, the
\textbf{distance matrix} $D(U,G)$ is defined as:
$(D(U,G))_{ij}=|i-j|$. $\diam(U)=\max\left(D(U,G)\right)$ is the
\textbf{diameter} of $U$ in $G$. For a given $d$, a distance matrix
is 'good', if it can be embedded in a $d-$regular tree. A subset
$U\subset V$ is good, if its distance matrix is good. We should note
that if $G$ is an $(n,d)$ graph, then (due to the local tree
property) almost any $U\subset V$ with $\diam(U)<\log n/2$ is good.\\
For $U\subset V$ and $\lambda\in[-2\sqrt{d-1},2\sqrt{d-1}]$, we
define the \textbf{limiting covariance matrix}
$\mathcal{C}_\lambda(U)$ by
$(\mathcal{C}_\lambda)_{ij}=\Cov_{D_{ij}}^{tree}(\lambda)$, where
$\Cov_{k}^{tree}(\lambda)$ is given by (\ref{eq:CovExp2}).\\
We will denote by $p(\mathbf{x},\Sigma)$ and
$\Phi(\mathbf{x},\Sigma)$, the density and the cdf of the
multinormal variable $\mathbf{x}=(x_1,...x_m)^T$ with mean zero and
covariance matrix $\Sigma$:
 \be\label{eq:MultiNormalDist}
  p(\mathbf{x},\Sigma)=\frac1{\sqrt{(2\pi)^m |\Sigma|}}
  e^{-\langle \mathbf{x},\Sigma^{-1}\mathbf{x} \rangle/2} \quad ,\quad
  \Phi(\mathbf{x},\Sigma)=\int_{-\infty}^\mathbf{x} p(\mathbf{y},\Sigma)d\mathbf{y}
 \ee
Equipped with this notation, we suggest the following for
$\lambda_0\in[-2\sqrt{d-1},2\sqrt{d-1}]$ and a fixed $m$:
 \begin{itemize}
 \item \textbf{Hypothesis I}(multivariate): For almost any $G(n,d)=(V,E)$ and a good
  $U\subset V$ of small diameter in $G$, the probability that $f(U)=(f(u_1),...f(u_m))^T<\mathbf{x}$,
  where $f$ is an adjacency eigenvector, chosen uniformly from the spectral window
  $[\lambda_0,\lambda_0+\epsilon]$, is bounded by
   \be\label{eq:densityGnd}
    \left|\mathbb{P}(f(U)<\mathbf{x})-\Phi(\mathbf{x},\mathcal{C}_{\lambda_0})\right|<\Delta(n,\epsilon)
   \ee
  where $\Delta(n,\epsilon)\rightarrow0$ as $\epsilon\rightarrow0$ but $\epsilon n\rightarrow\infty$.
 \item \textbf{Hypothesis II} (multivariate): For a good distance matrix $D$,
  the distribution of $f^{(\lambda_0)}(U)$ converges as $n\rightarrow\infty$, to
  the multinormal distribution (\ref{eq:MultiNormalDist}) with covariance matrix
  $\mathcal{C}_{\lambda_0}$, where $f^{(\lambda_0)}$ is an adjacency eigenvector
  with the closest eigenvalue to $\lambda_0$ of a uniformly chosen $(n,d)$ graph
  and $U$ satisfies $D_G(U)=D$.
\end{itemize}
Two remarks are in order. First, in hypothesis I we avoid the
question how small should $\diam(U)$ be, as we base the hypothesis
mainly on numerical simulations, which are applicable for
small-diameter distance matrices only (see section
\ref{subsec:NodalCount}). Second, we should note that in equation
(\ref{eq:MultiNormalDist}) we assume the existence of the inverse
covariance matrix. The existence of an inverse for
$\mathcal{C}_{\lambda_0}$ will be discussed in the next section,
where we show that for distance matrices which contain a vertex and
all of its neighbors, the covariance is singular. We also
demonstrate how this calculative obstacle can be removed by a simple
coordinate transformation.\\
Unlike the univariate conjectures, a comprehensive numerical
examination of the multivariate versions is a hard task. As a
beginning, we had to make do with the comparison of the empirical
cdf of two adjacent vertices to (\ref{eq:densityGnd}), where for
this configuration $\mathcal{C}_{\lambda}$ is given by
$(\mathcal{C}_\lambda)_{11}=(\mathcal{C}_\lambda)_{22}=1,
(\mathcal{C}_\lambda)_{12}=(\mathcal{C}_\lambda)_{21}=\lambda/d$.\\
For $k$ iid bivariate normal variables with mean zero and covariance
$\mathcal{C}_\lambda$, the empirical cdf is expected to converge to
$\Phi(\mathbf{x},\mathcal{C}_\lambda)$ as $1/\sqrt{k}$. In order to
check the second hypothesis we have measured the value of two
adjacent vertices $(f^{(j)}_1,f^{(j)}_2)$ over 20000 realizations of
$(6000,4)$ graphs from the eigenvector with the closest eigenvalue
to some $\lambda$. In order to evaluate the rate of convergence as a
function of the number of realizations, we have calculated for
various values of $k$, the empirical cdf for the first $k$
measurements:
 \be\label{eq:EmpComDis2}
  \mathcal{F}_\lambda^k(\mathbf{x})=\frac1k \{\#i|i\le k,f^{(j)}_1<x_1,f^{(j)}_2<x_2\}
 \ee
Finally, for every $lambda$ and $k$, we have calculated
$\max\left|\Phi(\mathbf{x},\mathcal{C}_\lambda)-
\mathcal{F}_\lambda^k(\mathbf{x})\right|$. As demonstrated in figure
\ref{fig:CheckTwo} the deviation does decrease as $k$ increases,
however the convergence is slower than the one measured for iid
bivariate normal variables. In addition, for larger eigenvalues
(smaller Laplacian eigenvalues), we have observed a faster
convergence.\\
The first hypothesis is harder to test directly, as it requires the
generating (and more problematic, the diagonalization) of a very
large graph, in order to have a narrow spectral window which
contains many eigenvectors. By using MATLAB's function eigs.m We
have explored relatively narrow spectral windows
($\lambda_{max}-\lambda_{min}\approx0.1$) of $(13000,3)$ graphs,
which contains between 200 to 400 eigenvectors (the exact number
depends on the spectral density at $\lambda$). The KS distance
between the empirical cdf for those windows and
$\Phi(\mathbf{x},\mathcal{C}_\lambda)$, was consistent with the
measured deviation in the previous experiment for the same number
of samples.\\
An additional support to the Gaussian approximation will be
introduced in the next section, where we reconstruct the structure
of the nodal pattern of an eigenvector, assuming the suggested
normal distribution.

\section{The nodal structure of eigenvectors}\label{sec:NodalStructure}
For a graph $G=(V,E)$ and a function $f(V)$, we define the
\textbf{induced nodal graph} $\tilde{G}_f=(V,\tilde{E}_f)$ , by the
deletion of edges, which connect vertices of opposite signs in $f$:
$\tilde{E}_f=\{(v_i,v_j)\in E| f_i f_j>0\}$. In this section we
analyze the nodal pattern of the eigenfunctions, \textbf{assuming
the multinormal distribution}, as stated in hypothesis II. We will
demonstrate that this assumption allows us not only to evaluate the
expectation of the nodal count, but also to estimate the
distribution of the size and shape of domains. In particular we will
demonstrate that the nodal structure cannot be imitated by
percolation-like models.

\subsection{Distribution of valency}\label{subsec:valency}
We begin by calculating $p_e(\lambda)$ - the probability of an edge
$e\in E$ of a random graph, to belong to $\tilde{E}_f$ for an
eigenvector $f$ with eigenvalue $\lambda$. This is twice the
probability of two adjacent vertices to be positive, which according
to hypothesis II, equals
 \be\label{eq:Pe}
  p_e(\lambda)=2\int_0^\infty\int_0^\infty d\mathbf{f} \frac1{2\pi\sqrt{|\mathcal{C}_\lambda|}}
  \exp\left(-\frac12 \langle\mathbf{f},\mathcal{C}_\lambda^{-1}\mathbf{f}\rangle\right)
 \ee
where $(\mathcal{C}_\lambda)_{11}=(\mathcal{C}_\lambda)_{22}=1$ ,
$(\mathcal{C}_\lambda)_{12}=(\mathcal{C}_\lambda)_{21}=\lambda/d$.
Integrating, we get that:
 \be\label{eq:Pe2}
  p_e(\lambda)=\frac12+\frac1\pi\arcsin\left(\frac{\lambda}d\right)
 \ee
\begin{figure}[h]
\centering
 \scalebox{0.6}{\includegraphics{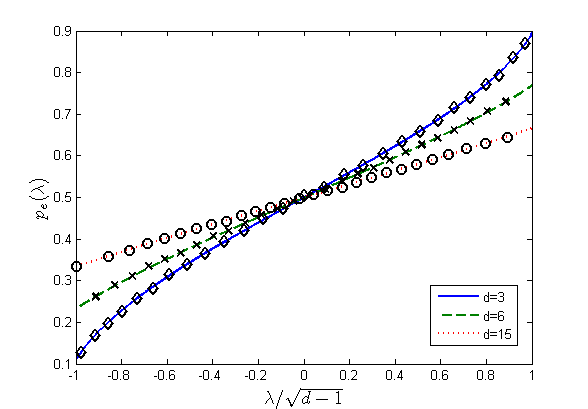}}
  \caption{A comparison between the Gaussian prediction for
  $p_e(\lambda)$ (lines) and the empirical result (markers) for a single
  random realization of 3,6 and 15 regular graphs on 4000 vertices.}
    \label{fig:p_e}
\end{figure}
$p_e(\lambda)$ is symmetric with respect to $\lambda=0$. In
addition, since $|\lambda|\le2\sqrt{d-1}$, for small values of $d$,
$p_e(\lambda)$ varies considerably along the spectrum (thus, for
$d=3$, $p_e(\lambda)$ can take values in the interval $[0.1,0.9]$),
while for large $d$ the changes are moderate (for $d=100$ for
example, it is constrained to $[0.44,0.56]$). As demonstrated in
figure \ref{fig:p_e}, this result describes with high accuracy the
observed probability, for various values of $d$.\\
The Gaussian model predicts as well a distribution for the valency
of vertices in $\tilde{G}_f$. In order to evaluate $p_j(\lambda)$ -
the probability of a vertex $v_0\in V$ to be of valency $j$ in
$\tilde{G}_f$, one should consider the mutual distribution of $f_0$
and its $d$ neighbors $\{f_1,...f_d\}$. The appropriate covariance
entries are given by
$(\mathcal{C}_\lambda)_{ii}=1,(\mathcal{C}_\lambda)_{0j}=
(\mathcal{C}_\lambda)_{j0}=\lambda/d$ and
$(\mathcal{C}_\lambda)_{jk}=(\lambda^2-d)/d(d-1)$, for $0\le i\le d$
,$1\le j,k\le d$ and $j\ne k$.\\
This matrix is singular, as $(\lambda,-1,-1,...-1)^T$ is an
eigenvector of $\mathcal{C}_\lambda$ with zero eigenvalue. This
singularity is due to the constraint $\lambda f_0=\sum_j f_j$ which
is kept by the Gaussian model. In order to avoid the singularity we
may integrate with respect to the new Gaussian variables
$\tilde{\mathbf{f}}=(\lambda f_0-\sum_j f_j,f_1,...f_d)^T$.
Introducing the (invertible) matrix $\tilde{\mathcal{C}}_\lambda$,
obtained from $\mathcal{C}_\lambda$ by changing the of diagonal
terms in the zeroth row and column to zero, $p_j(\lambda)$ can be
written as:
 \be\label{eq:Pj}\fl
  p_j(\lambda)=2{d\choose j}\int \frac{d\tilde{\mathbf{f}}}{\sqrt{(2\pi)^d|\tilde{\mathcal{C}}_\lambda|}}
  \delta(\tilde{f}_0)\theta(f_0)\prod_{i=1}^j\theta(f_i)\prod_{l=j+1}^d\theta(f_l)
  \exp[-\frac12\langle\tilde\mathbf{f},\tilde{\mathcal{C}}_\lambda^{-1}\tilde\mathbf{f}\rangle]
 \ee
\begin{figure}[h]
\centering
 \scalebox{0.7}{\includegraphics{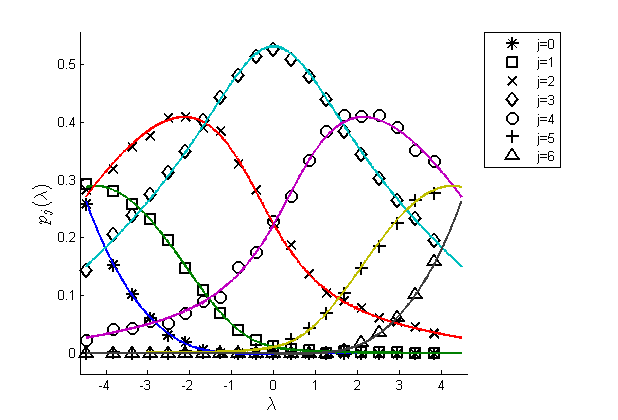}}
   \caption{A comparison between the Gaussian prediction of
  $p_j(\lambda)$ and the empirical result for a single realization of
  $G(4000,6)$.}
    \label{fig:p_j}
\end{figure}
where the prefactor is due to the sign symmetry and the different
alternatives to choose $j$ out of $d$ adjacent vertices. An
immediate result of this expression is the symmetry
$p_j(\lambda)=p_{d-j}(-\lambda)$. This integral cannot be calculated
explicitly, however it can be evaluated, e.g. by the method of
\cite{Genz}. As in the study of $p_e(\lambda)$ the Gaussian
prediction is very close to the observed results. As an example, in
figure \ref{fig:p_j} we compare the Gaussian prediction for
$p_j(\lambda)$ and $d=6$, evaluated by the function qscmvnv.m
\cite{GenzCode}, to the measured result for a single realization of
a $(4000,6)$ graph.

\subsection{The nodal count of an eigenfunction}\label{subsec:NodalCount}
\begin{figure}[h]
\centering
 \scalebox{0.7}{\includegraphics{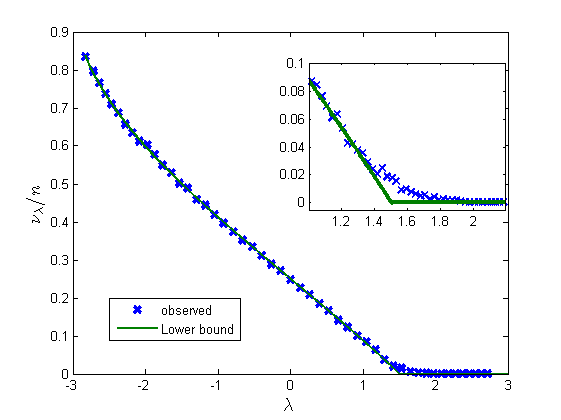}}
  \caption{A comparison between (\ref{eq:LowBound1}) and the
  observed count for a single realization of $G(4000,3)$. The
  inset is a magnification of the spectral window near the
  bound's flexion - the only part of the spectrum in which the
  observed count deviates considerably from the bound.}
    \label{fig:nc3_4000}
\end{figure}

\begin{figure}[h]
\centering
 \scalebox{0.6}{\includegraphics{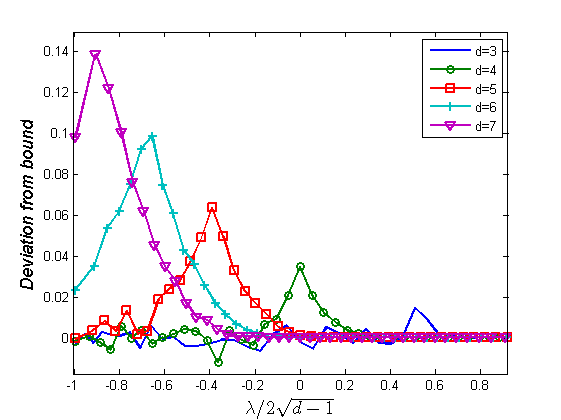}}
     \caption{The deviation of $\nu_\lambda/n$ from the lower bound
     (\ref{eq:LowBound1}), for 3,4,5,6 and 7 regular graphs on 4000
     vertices.} \label{fig:nc34567_4000}
\end{figure}

\begin{figure}[h]
\centering
 \scalebox{0.6}{\includegraphics{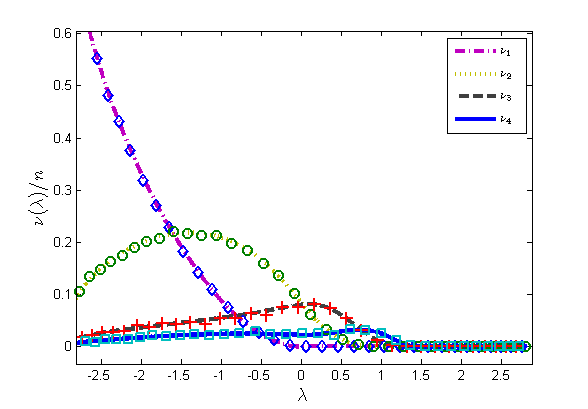}}
     \caption{The Gaussian prediction for
     $\nu_k/n$ for $d=3$ and $1\le k\le40$ (lines), compared to the observed count
     (markers) for a single realization of $G(4000,3)$.}
    \label{fig:nu3}
\end{figure}

\begin{figure}[h]
\centering
 \scalebox{0.5}{\includegraphics{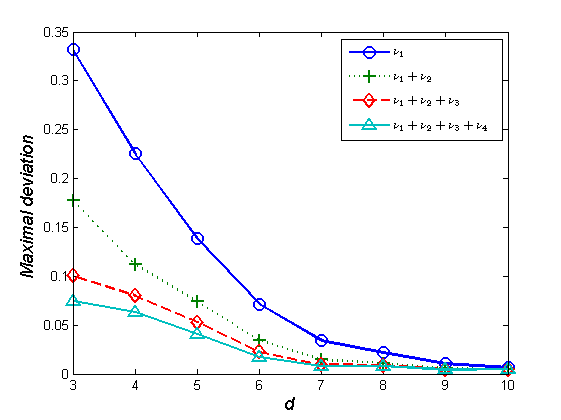}}
     \caption{The maximal deviation between the measured nodal count
     of $3\le d\le10$ regular graphs on 4000 vertices to the
     expected number of domains, smaller or equal than 1,2,3 and 4
     vertices.}
    \label{fig:Cdev}
\end{figure}

In \cite{Dekel}, the following intriguing properties of the nodal
count $\{\nu_j\}_{j=1}^n$ for the eigenvectors of $(n,d)$ graphs
were observed. First, for all $j<j_0(d,n)$, $\nu_j$ was found to be
exactly $2$, where the relative part $j_0/n$ of eigenvectors with
exactly two nodal domains is increasing with $d$. Second, for small
values of $d$, and for $j>j_0$, the nodal count increases
approximately linearly with $j$. While the known bounds on the nodal
count (see section \ref{subsec:GraphsIntro}) are far from being
satisfactory in explaining this behavior, we would like to
demonstrate in this section, how does the expected nodal count
emerges from the Gaussian model.\\
Adopting the Gaussian expression (\ref{eq:Pe2}) for $p_e(\lambda)$,
it is possible to derive a lower bound on the expected nodal count
of an eigenvector. The number $N$ of connected components of a graph
$G=(V,E)$, is given by $N=V-E+C$, where $C$ is the number of
independent cycles in $G$. Since on average, the induced nodal graph
$\tilde{G}_f$ posses $p_e(\lambda)\cdot|E|=\frac{nd}2 p_e(\lambda)$
edges and $n$ vertices, the expected nodal count is bounded from
below (for all of the eigenvectors but the first) by
 \be\label{eq:LowBound1}
  E\left(\nu(n,\lambda)\right)\ge\max\left\{2,n\left(1-\frac{d}2 p_e(\lambda)\right)\right\}
 \ee
We should note that this bound is effective only for $d\le7$, as for
larger values, $1-dp_e(\lambda)/2$ is negative for
$|\lambda|\le2\sqrt{d-1}$.\\
For low values of $d$ this crude bound matches surprisingly well the
observed nodal count, as is demonstrated in figure
\ref{fig:nc3_4000} for a $(4000,3)$ graph.
The good agreement can be understood if we consider the critical
properties of the nodal pattern. Numerical observations
\cite{ElonIP} suggest that for $d\le5$, the induced nodal graph
$\tilde{G}_f$ exhibits a phase transition at some $\lambda_c$ as
$n\rightarrow\infty$. In the subcritical phase ($\lambda<
\lambda_c$), the size of the largest nodal domains is proportional
to $\log n$, while in the supercritical phase ($\lambda>\lambda_c$),
two giant components of order $n$ emerge.\\
As the number of connected components of size $\log n$ in an $(n,d)$
graph, which contain $\log(\log n)$ cycles is almost surely zero,
the expected number of independent closed cycles in $\tilde{G}_f$ in
the subcritical regime must be much smaller than $n\log(\log n)/\log
n$ (as there cannot be more than $n/\log n$ domains comparable in
size to $\log n$). As a result, for $\lambda<\lambda_c$ the
deviation between (\ref{eq:LowBound1}) to the expected count is at
most of order $1/\log(n)$. This result is reflected in figure
\ref{fig:nc34567_4000}, which demonstrate that for $d\le 5$ (where a
subcritical phase is observed), the measured count converges to
(\ref{eq:LowBound1}), for low enough eigenvalues.\\
The fact that only two nodal domains are observed for a large number
of eigenvectors is also consistent with the existence of a
supercritical phase. A general property of supercritical systems is
the scarcity of large but finite clusters. the expected number of
clusters of size $s$ decays asymptotically as
$\exp(-s(p-p_c)^\gamma)$, for some (model dependent) positive
$\gamma$. Therefore, the supercritical phase consists of a giant
component and 'dust'. When we consider the nodal pattern of
supercritical eigenfunction (i.e. those with $\lambda>\lambda_c$)
two special phenomena occur. The first is the appearance of two
giant components - a positive and a negative domains. The second is
the rarity of small domains: as was mentioned in section
\ref{subsec:GraphsIntro}, the distribution of the eigenvectors is
constrained, preventing the existence of small domains for large
enough values of $\lambda$. As a result, we expect to find only
rarely more than 2 nodal domains, for a considerable amount of first
eigenvectors (which are deep enough in the supercritical regime).
Moreover, as $d$ increases, the value of $\lambda_c$ decreases,
therefore the expected number of such eigenvectors is supposed to
increase with $d$ (as is indeed observed).\\
As the size distribution of clusters is expected to decay rapidly,
we can tighten the bound on the expected nodal count considerably,
by calculating $\nu_k(n,\lambda)$ - the expected number of domains
of size $k$ for small values of $k$. It is easy to see that
$\nu_1(n,\lambda)=np_0(\lambda)$ where $p_0(\lambda)$ is given by
(\ref{eq:Pj}). For $k>1$ the calculation can be carried out in the
same spirit: The probability for $k$ given vertices to form a nodal
domain of size $k$ can be evaluated, through a $(k(d-1)+2)$
dimensional integral (over the $k$ vertices and their $k(d-2)+2$
neighbors), in a similar manner to (\ref{eq:Pj}). Finally,
$\nu_k(n,\lambda)$ is given (up to small corrections) by summing the
probabilities over all trees of size $k$ and maximal valency $d$,
multiplied by the number of such trees in $G$. The agreement between
the Gaussian prediction to the observed distribution of $\nu_k$ is
demonstrated in figure \ref{fig:nu3}.
As $E(\nu)=\sum_{k=1}^\infty \nu_k$, we get that
$\sum_{i=1}^k\nu_k(n,\lambda)$ should converge to
$E(\nu(\lambda,n))$. In figure \ref{fig:Cdev} we plot the maximal
deviation between $\sum\nu_k$ (for $1\le k\le4$) and the measured
count of $(4000,d)$ graphs for $3\le d\le10$. It can be seen that
the converges is much faster for $d>5$. This is consistent with
relatively slow decay in the size distribution, which is expected in
the vicinity of the critical point.

\subsection{Eigenvectors and percolation}
For a $G(n,d)$ graph and $0\le p\le1$, the induced percolation graph
$G(n,d,p)$ is obtained by deleting the edges of $G$ independently
with probability $1-p$. As was mentioned in section \ref{subsec:QC},
for two-dimensional billiards, it is believed that the nodal pattern
exhibits (in the semi-classical limit) a percolation-like behavior
\cite{Bogomolny02}. In this section we compare the properties of
$G(n,d,p)$ to the nodal pattern of an eigenvector, satisfying
$p_e(\lambda)=p$ (see eq. \ref{eq:Pe2}).\\
An important difference between the two models is their valency
distribution. For percolation, as the deletion of different edges is
independent, the probability of a vertex in $G(n,d,p)$ to be of
valency $j$ is given by
 \be\label{eq:Pperc}
  p_j^{perc}(p)={d\choose l}p^l(1-p)^{d-l}
 \ee
which is essentially different from $p_j(\lambda)$ (\ref{eq:Pj}) -
the equivalent expression for the nodal pattern. As will be
demonstrated soon, by changing the valency distribution, we change
global properties of the pattern as well.\\
The differences between the two processes are not limited to local
measures such as $p_j$, but also for events which involve several
vertices. For example, the probability to have a connected cluster
of size $k$ in $G(n,d,p)$ is positive for any $p>0$, and can be
expressed through $p_j^{perc}(p)$. As for the nodal pattern, for any
$k\in\mathbb{N}$ there is some $\lambda_{max}$, so that for
$\lambda>\lambda_{max}$ the probability to have a domain smaller
than $k$ is zero.\footnote{This is so, as the probability to find a
connected component in $G$  of size $k$ with more than 1 cycle goes
to zero as $n\rightarrow\infty$, allowing the use of bounds
reminiscent of \cite{Leydold}.} In addition, as was demonstrated
above, the quantities $\nu_k$ cannot be reduced to simple functions
of $p_j(\lambda)$.\\
Finally, we would like to show that the critical threshold for the
two models is different. In \cite{Alon04} the critical probability
for $G(n,d,p)$ was found to be $p_c=1/(d-1)$. It is not hard to
verify that for the nodal pattern $p_e(\lambda_c)>1/(d-1)$. Consider
for example the case $d=3$, where $p_c=1/2=p_e(\lambda=0)$. As was
mentioned before, for $\lambda=0$ there are no interior points in
the nodal domains, therefore, for $d=3$ all nodal domains must be
linear chains. however, percolation on linear chains is always
subcritical, therefore necessarily $\lambda_c>0$. By similar
arguments, this will also be the case for $d>3$.\\
We would like to note that the mismatch between the properties of
the Laplacian nodal pattern and $G(n,d,p)$ should not come as a
great surprise. One of the main arguments in favor of the
percolation model \cite{Bogomolny07} for 2 dimensional billiards, is
that the asymptotic rate of decay of the eigenvectors' covariance
(\ref{eq:CovBerry2}) is fast enough in order to neglect
correlations, according to the so called 'Harris criterion'
\cite{Harris}. However, the covariance for eigenvectors of $(n,d)$
graphs (\ref{eq:CovExp2}), does not fulfill the requirements of this
criterion, implying that the scaling limit of the two models will
differ. We should note that the requirements of this criterion are
not fulfilled for billiards in more than two dimensions as well.
This suggest that the resemblance between the nodal pattern of
Laplacian eigenvectors to percolation is a two dimensional
phenomenon.

\section{Conclusions}
As a summary, we collect the main new results of this work
concerning the structure of Laplacian (Adjacency) eigenvectors:
\begin{enumerate}
 \item
  In the limit $n\rightarrow\infty$, the empirical covariance (\ref{eq:covDef}) of
  an eigenvector of a uniformly chosen $(n,d)$ graph, for a distance
  $k<log_{d-1}(n)/2$ is given by
   \be
    \fl\quad\Cov_k^{tree}(\lambda)=\frac1{d(d-1)^{k/2}}
    \left((d-1)U_k\left(\frac{\lambda}{2\sqrt{d-1}}\right)-
    U_{k-2}\left(\frac{\lambda}{2\sqrt{d-1}}\right)\right)
   \ee
  which decays exponentially with $k$. For $k>\log_{d-1}(n)/2$, this approximation
  loses its accuracy. however, the observed rate of decay is still
  exponential in $k$.
 \item
  We provide numerical evidence in support of the hypothesis that the distribution
  of the adjacency (or Laplacian) eigenvectors follows the Gaussian measure
   \be
    p(f(U))=\frac1{\sqrt{(2\pi)^m|\mathcal{C}_{\lambda_0}|}}
    \exp\left[-\frac12 \langle f(U), \mathcal{C}_{\lambda_0}^{-1}f(U)\rangle\right]
   \ee
  For any $\lambda_0\in[-2\sqrt{d-1},2\sqrt{d-1}]$.
 \item
  We used the Gaussian measure to predict the expected number of
  nodal domains in an eigenfunction and its dependence in the
  eigenvalue. We have shown the consistency of the Gaussian hypothesis
  with various nodal properties, such as valency and size distribution.
 \item
  We have shown that the nodal structure of the Laplacian eigenvectors differ
  from the cluster structure of $G(n,d,p)$. The two models are not sharing
  the same critical point, and the structure of a typical components does not
  follow the same law in the two cases.
\end{enumerate}
\ack {I wish to thank U. Smilansky for guiding me patiently through
this work - without his help and advises, this paper could have not
been written. I am grateful to N. Linial for exposing me to the
fascinating world of expanders, and for raising some of the problems
I came to investigate. I would like to acknowledge I. Oren, G.
Kozma, O. Zeitouni, Y. Rinott, S. Sodin, J. Breuer and A. Aronovitch
for useful discussions and remarks. Thanks to A. Genz for sharing
his codes, and adjusting it to our needs. The work was supported by
the Minerva Center for non-linear Physics and the Einstein (Minerva)
Center at the Weizmann Institute, and by grants from the BSF (grant
2006065), the GIF (grant I-808-228.14Q2003) and the ISF (grant
168/06).}

\noindent {\bf{Bibliography}}
\bibliographystyle{unsrt}
\bibliography{geometric}

\begin{thebibliography}{10}

\bibitem{Dekel}
Lee~J Dekel~Y and Linial N.
\newblock chapter Eigenvectors of Random Graphs: Nodal Domains, pages 436--448.
\newblock 2007.
\newblock 10.1007/978-3-540-74208-1\_32.

\bibitem{Bogomolny02}
Bogomolny E and Schmit C.
\newblock {Percolation Model for Nodal Domains of Chaotic Wave Functions}.
\newblock {\em Physical Review Letters}, 88(11):114102, March 2002.

\bibitem{LinialExpanders}
Linial~N Hoory~S and Wigderson A.
\newblock Expander graphs and their applications.
\newblock {\em Bull. Amer. Math. Soc. (N.S.)}, 43(4):439--561 (electronic),
  2006.

\bibitem{Shi}
Shi J and Malik J.
\newblock Normalized cuts and image segmentation.
\newblock {\em IEEE Transactions on Pattern Analysis and Machine Intelligence},
  22(8):888--905, 2000.

\bibitem{Coifman}
Coifman~R R.
\newblock Perspectives and challenges to harmonic analysis and geometry in high
  dimensions: geometric diffusions as a tool for harmonic analysis and
  structure definition of data.
\newblock In {\em Perspectives in analysis}, volume~27 of {\em Math. Phys.
  Stud.}, pages 27--35. Springer, Berlin, 2005.

\bibitem{Pothen}
Simon H~D Pothen~A and Liou~K P.
\newblock Partitioning sparse matrices with eigenvectors of graphs.
\newblock {\em SIAM J. Matrix Anal. Appl.}, 11(3):430--452, 1990.
\newblock Sparse matrices (Gleneden Beach, OR, 1989).

\bibitem{Lovasz}
Lov{\'a}sz L.
\newblock Random walks on graphs: a survey.
\newblock In {\em Combinatorics, Paul Erd\H os is eighty, Vol.\ 2 (Keszthely,
  1993)}, volume~2 of {\em Bolyai Soc. Math. Stud.}, pages 353--397. J\'anos
  Bolyai Math. Soc., Budapest, 1996.

\bibitem{Rudnick}
Rivin~I Jakobson~D, Miller S~D and Rudnick Z.
\newblock {Eigenvalue spacings for regular graphs}.
\newblock {\em ArXiv High Energy Physics - Theory e-prints}, September 2003.

\bibitem{Smilansky07}
Smilansky U.
\newblock Quantum chaos on discrete graphs.
\newblock {\em Journal of Physics A: Mathematical and Theoretical},
  40(27):F621--F630, 2007.

\bibitem{Berry77}
Berry~M V.
\newblock Regular and irregular semiclassical wave functions.
\newblock {\em Journal of Physics A Mathematical General}, 10:2083--2091, 1977.

\bibitem{Urbina}
Urbina~J D and Richter K.
\newblock {Semiclassical construction of random wave functions for confined
  systems}.
\newblock {\em Physical Review E}, 70(1):015201--+, July 2004.

\bibitem{Blum}
Gnutzmann~S Blum~G and Smilansky U.
\newblock Nodal domains statistics: A criterion for quantum chaos.
\newblock {\em Physical Review Letters}, 88(11):114101, March 2002.

\bibitem{Elon}
Joas~C Elon~Y, Gnutzmann~S and Smilansky U.
\newblock Geometric characterization of nodal domains: the area-to-perimeter
  ratio.
\newblock {\em J. Phys. A}, 40(11):2689--2707, 2007.

\bibitem{Courant}
Courant R and Hilbert D.
\newblock {\em Methods of mathematical physics. {V}ol. {I}}.
\newblock Interscience Publishers, Inc., New York, N.Y., 1953.

\bibitem{Foltin04}
Gnutzmann~S Foltin, G and Smilansky U.
\newblock The morphology of nodal lines random waves versus percolation.
\newblock {\em Journal of Physics A Mathematical General}, 37:11363--11371,
  November 2004.

\bibitem{Aronovitch}
Aronovitch A and Smilansky U.
\newblock {The statistics of the points where nodal lines intersect a reference
  curve}.
\newblock {\em Journal of Physics A Mathematical General}, 40:9743--9770,
  August 2007.

\bibitem{Bogomolny07}
Bogomolny E and Schmit C.
\newblock {Random wavefunctions and percolation}.
\newblock {\em Journal of Physics A Mathematical General}, 40:14033--14043,
  November 2007.

\bibitem{BogomolnySle}
Dubertrand~R Bogomolny~E and Schmit C.
\newblock {SLE description of the nodal lines of random wave functions}.
\newblock {\em ArXiv Nonlinear Sciences e-prints}, September 2006.

\bibitem{Keating}
Marklof~J Keating J~P and Williams~I G.
\newblock Nodal domain statistics for quantum maps, percolation, and stochastic
  loewner evolution.
\newblock {\em Physical Review Letters}, 97(3):034101, 2006.

\bibitem{Bollobas80}
Bollob{\'a}s B.
\newblock A probabilistic proof of an asymptotic formula for the number of
  labelled regular graphs.
\newblock {\em European J. Combin.}, 1(4):311--316, 1980.

\bibitem{Bollobas}
Bollob\'as B and De~la~Vega W~F.
\newblock The diameter of random regular graphs.
\newblock {\em Combinatorica}, 2(2):125--134, June 1982.

\bibitem{McKay}
McKay~B D.
\newblock The expected eigenvalue distribution of a large regular graph.
\newblock {\em Linear Algebra Appl.}, 40:203--216, 1981.

\bibitem{Davies}
Leydold~J Davies E~B, Gladwell G M~L and Stadler~P F.
\newblock Discrete nodal domain theorems.
\newblock {\em Linear Algebra Appl.}, 336:51--60, 2001.

\bibitem{Band}
Oren~I Band~R and Smilansky U.
\newblock Nodal domains on graphs - how to count them and why?, 2007.

\bibitem{Leydold}
B{\i}y{\i}ko{\u{g}}lu T and Leydold J.
\newblock Faber-{K}rahn type inequalities for trees.
\newblock {\em J. Combin. Theory Ser. B}, 97(2):159--174, 2007.

\bibitem{Biyikouglu}
Leydold~J B{\i}y{\i}ko{\u{g}}lu~T and Stadler~P F.
\newblock {\em Laplacian eigenvectors of graphs}, volume 1915 of {\em Lecture
  Notes in Mathematics}.
\newblock Springer, Berlin, 2007.
\newblock Perron-Frobenius and Faber-Krahn type theorems.

\bibitem{Alon06}
Lubetzky~E Alon~N, Benjamini~I and Sodin S.
\newblock Non-backtracking random walks mix faster, 2006.

\bibitem{Abramowitz}
Abramowitz M and Stegun~I A.
\newblock {\em Handbook of Mathematical Functions with Formulas, Graphs, and
  Mathematical Tables}.
\newblock Dover, New York, ninth dover printing, tenth gpo printing edition,
  1964.

\bibitem{Steger}
Steger A and Wormald~N C.
\newblock Generating random regular graphs quickly.
\newblock {\em Combin. Probab. Comput.}, 8(4):377--396, 1999.
\newblock Random graphs and combinatorial structures (Oberwolfach, 1997).

\bibitem{Genz}
Genz A.
\newblock Numerical computation of multivariate normal probabilities.
\newblock {\em J. Comput. Graph. Statist.}, 1:141--150, 1992.

\bibitem{GenzCode}
Genz A.
\newblock {http://www.math.wsu.edu/faculty/genz/homepage}.

\bibitem{ElonIP}
Y~Elon and U~Smilansky.
\newblock Level sets of eigenfunctions on regular graphs.
\newblock {\em In preparation}.

\bibitem{Alon04}
Benjamini~I Alon, N and Stacey A.
\newblock Percolation on finite graphs and isoperimetric inequalities.
\newblock {\em Ann. Probab.}, 32(3A):1727--1745, 2004.

\bibitem{Harris}
Harris~A B.
\newblock Effect of random defects on the critical behaviour of ising models.
\newblock {\em Journal of Physics C: Solid State Physics}, 7(9):1671--1692,
  1974.

\end{thebibliography}
\end{document}